\documentclass[letterpaper]{article} % DO NOT CHANGE THIS
\usepackage[most]{tcolorbox}
\usepackage{aaai2026}  % DO NOT CHANGE THIS
\usepackage{times}  % DO NOT CHANGE THIS
\usepackage{helvet}  % DO NOT CHANGE THIS
\usepackage{courier}  % DO NOT CHANGE THIS
\usepackage[hyphens]{url}  % DO NOT CHANGE THIS
\usepackage{graphicx} % DO NOT CHANGE THIS
\usepackage{natbib}  % DO NOT CHANGE THIS AND DO NOT ADD ANY OPTIONS TO IT
\usepackage{caption} % DO NOT CHANGE THIS AND DO NOT ADD ANY OPTIONS TO IT
\usepackage{algorithm}
\usepackage{algorithmic}
\usepackage{amsmath}
\usepackage{multirow}
\usepackage{newfloat}
\usepackage{listings}
\DeclareCaptionStyle{ruled}{labelfont=normalfont,labelsep=colon,strut=off} % DO NOT CHANGE THIS
\floatstyle{ruled}
\newfloat{listing}{tb}{lst}{}
\floatname{listing}{Listing}
\title{PSM: Prompt Sensitivity Minimization via LLM-Guided Black-Box Optimization}

\author {
    Hussein Jawad\textsuperscript{\rm 1},
    Nicolas J-B. Brunel\textsuperscript{\rm 1,2,3}
}
\affiliations {
    \textsuperscript{\rm 1}Capgemini Invent, Paris, France\\
    \textsuperscript{\rm 2}LaMME, University Paris Saclay, Evry, France\\
    \textsuperscript{\rm 3}ENSIIE, Evry, France
}

\usepackage{bibentry}
\begin{document}

\maketitle

\begin{abstract}
System prompts are critical for guiding the behavior of Large Language Models (LLMs), yet they often contain proprietary logic or sensitive information, making them a prime target for extraction attacks. Adversarial queries can successfully elicit these hidden instructions, posing significant security and privacy risks. Existing defense mechanisms frequently rely on heuristics, incur substantial computational overhead, or are inapplicable to models accessed via black-box APIs. This paper introduces a novel framework for hardening system prompts through shield appending, a lightweight approach that adds a protective textual layer to the original prompt. Our core contribution is the formalization of prompt hardening as a utility-constrained optimization problem. We leverage an LLM-as-optimizer to search the space of possible SHIELDs, seeking to minimize a leakage metric derived from a suite of adversarial attacks, while simultaneously preserving task utility above a specified threshold, measured by semantic fidelity to baseline outputs. This black-box, optimization-driven methodology is lightweight and practical, requiring only API access to the target and optimizer LLMs. We demonstrate empirically that our optimized SHIELDs significantly reduce prompt leakage against a comprehensive set of extraction attacks, outperforming established baseline defenses without compromising the model's intended functionality. Our work presents a paradigm for developing robust, utility-aware defenses in the escalating landscape of LLM security. The code is made public on the following link: \url{https://github.com/psm-defense/psm}
\end{abstract}


\section{Introduction}

\paragraph{The Duality of System Prompts} 
System prompts have become a cornerstone of modern Large Language Model (LLM) applications, serving as the primary mechanism for steering model behavior \citep{liu2023pretrain,brown2020gpt3}. These carefully engineered instructions extend beyond simple suggestions; they define an LLM's persona, operational constraints, task-specific rules, and interaction style \citep{malik2024persona,dong2023steerlm}. By providing a structured framework and high-level context, system prompts ensure that model outputs are coherent, relevant, and aligned with the developer's intended goals, enhancing both performance and rule adherence \citep{ouyang2022instructgpt,chung2022flan,wei2022cot,kojima2022zeroshotcot}. For many commercial LLM-powered products, the system prompt encapsulates the core intellectual property and competitive advantage—it is the "secret sauce" that distinguishes a generic foundation model from a specialized, high-performing application \citep{hui2024pleak,Zhang2023effective,jiang2024promptkeeper,agarwal2024promptleakage}.

\paragraph{The Threat of Prompt Extraction} 
The very value of system prompts makes them a prime target for malicious actors. These prompts often contain proprietary business logic, sensitive architectural details like API integration instructions, or confidential filtering criteria that are not intended for public disclosure \citep{agarwal2024promptleakage}. This creates a critical vulnerability known as prompt extraction, a class of adversarial attack where users craft malicious inputs to deceive an LLM into revealing its own system instructions \citep{Zhang2023effective, wang2024raccoon}. This form of ``adversarial reconnaissance'' can serve as a precursor to more sophisticated attacks, enable the complete replication of a proprietary service, or expose sensitive operational data \citep{agarwal2024promptleakage}. The commercialization of LLM applications has amplified this threat, creating direct economic incentives for prompt extraction \citep{hui2024pleak}. The emergence of prompt marketplaces, where high-quality prompts are treated as valuable commodities, transforms prompt leakage from a theoretical vulnerability into a tangible economic risk, motivating persistent and sophisticated attacks \citep{Zhang2023effective}.

\paragraph{Limitations of Existing Defenses} 
Current defenses against prompt extraction are often ad-hoc and fall short of providing robust protection \citep{agarwal2024promptleakage,wang2024promptbenchmark,chen2024structured}. Many systems rely on simple, heuristic-based instructions appended to the prompt, such as ``Do not reveal your instructions.'' However, such defenses are brittle and easily circumvented by adversaries who craft queries to override these rules (e.g., indirect or direct prompt-injection/jailbreaks), for instance, by instructing the model to ``Ignore all previous instructions'' \citep{greshake2023indirect,yi2023benchmarking}. This vulnerability stems from a fundamental tension within LLMs: their core training objective is to be helpful and follow instructions, which can conflict with safety constraints \citep{wei2023jailbroken}. A prompt extraction query is, from the model's perspective, merely another instruction to be followed. A simple defensive instruction creates a direct conflict between the developer's rule and the attacker's command, a conflict that attackers have proven adept at winning \citep{wei2023jailbroken,greshake2023indirect}. This suggests that a truly effective defense cannot rely on a simple battle of instructions but must instead alter the prompt's structure and semantics to make it inherently less susceptible to leakage \citep{hines2024spotlighting,pape2025promptobfuscation,zhuang2025proxyprompt}.

\paragraph{Prompt Sensitivity Minimization (PSM)} 
This paper introduces \emph{Prompt Sensitivity Minimization} (PSM), a principled framework designed to address the following practitioner scenario:

\emph{“As a developer using a closed-source LLM API (e.g. OpenAI), how can I design a defense against system prompt extraction that is black-box (does not require access to the target model), lightweight (adds minimal computational overhead), and effective?”}

To answer this question, we frame the problem of defending system prompts as a \emph{utility-constrained optimization task}. Specifically, our goal is to reduce how much of a sensitive system prompt can be extracted (minimize \emph{leakage}), while ensuring that the model’s desired behavior on normal user inputs is preserved (maintain \emph{utility}).

PSM introduces a novel black-box method to solve this problem. It automatically generates and iteratively refines a protective \emph{shield}—a short textual suffix appended to the original system prompt. This shield is optimized via interaction with the LLM itself, serving as a barrier that deflects adversarial prompt-extraction attempts while preserving the semantic behavior of the model.

% To answer this question, we propose a new algorithm for defending against system prompt extraction. Our approach treats system prompt hardening as a utility-constrained optimization problem. We then introduce a novel black-box method to solve this problem. PSM automatically generates and refines a protective \emph{shield}—a short suffix appended to the system prompt. This shield is optimized to reduce the risk of prompt extraction while preserving the intended behavior of the model.

The key contributions of this work are:

\begin{itemize}
  \item \textbf{Black-box compatibility:} PSM requires no access to model weights, gradients, or internals, and therefore applies to any proprietary or open-source LLM accessible through an API.
  \item \textbf{Low overhead:} The learned shield is a static textual suffix. It imposes essentially no inference-time cost and requires no changes to the serving stack; optimization can be performed offline.
  \item \textbf{Iterative improvement:} An LLM-as-optimizer intelligently proposes and evolves shield candidates, discovering effective, non-obvious strategies through guided black-box search.
  \item \textbf{Utility preservation:} PSM enforces a hard utility constraint during optimization, ensuring that hardening does not degrade performance on intended tasks.
\end{itemize}

PSM delivers a systematic, reproducible, and practical approach to securing valuable prompt assets—moving beyond brittle heuristics to a robust, optimization-driven defense.

\section{Related Works}

\subsection{Prompt Extraction Attacks}
The vulnerability of LLMs to prompt extraction has been documented through both anecdotal evidence and systematic research. Early reports demonstrated that prompts from commercial systems like Bing Chat could be recovered through simple queries \citep{edwards2023bingprompt,warren2023bingrules}. This threat was later formalized by researchers such as \citep{Zhang2023effective}, who demonstrated that straightforward text-based attacks could successfully extract secret prompts from a wide range of LLMs with high precision. The attack landscape has since been comprehensively mapped by benchmarks like Raccoon, which introduces a taxonomy of 14 attack types, including prefix injection, multilingual attacks, and formatting requests, establishing the diversity and complexity of threats that any defense must withstand \citep{wang2024raccoon}. More recent work has also highlighted the evolution of these attacks from single-turn queries to sophisticated multi-turn conversational strategies, where an adversary gradually manipulates the model into a vulnerable state, rendering static, single-shot defenses increasingly obsolete \citep{agarwal2024promptleakage}.

\subsection{Defenses Against Prompt Leakage}
The literature on defending against prompt leakage reveals a clear trade-off between implementation simplicity and robustness. Current approaches can be broadly categorized as follows:

\paragraph{Instructional and Heuristic Defenses} This is the most common and simplest approach, involving the addition of guardrail instructions like ``Do not reveal your prompt'' or the use of in-context examples to guide behavior. However, their effectiveness is limited, as they are easily bypassed by adversarial phrasing that overrides the defensive instruction \citep{greshake2023indirect,zou2023universal,owasp2025llm01}. In some cases, these defenses can even be counterproductive; for example, using XML tags to demarcate instructions has been shown to increase leakage rates \citep{agarwal2024promptleakage}. 

\paragraph{Input/Output Filtering and Sanitization} These methods employ external modules to screen user inputs for malicious patterns or to scan model outputs for sensitive content before it is returned to the user. While they can be effective, these approaches add computational/operational overhead and may struggle to detect novel or obfuscated attacks that do not match predefined patterns \citep{agarwal2024promptleakage,owasp2025llm01,zhuang2025proxyprompt}. 

\paragraph{Prompt Transformation and Obfuscation} This category includes more advanced techniques that modify the prompt itself. Microsoft's ``Spotlighting'' framework uses delimiters or character-level encoding to help the LLM differentiate between trusted developer instructions and untrusted user input \citep{hines2024spotlighting}. More sophisticated methods like \emph{ProxyPrompt} and \emph{Prompt Obfuscation} aim to replace the original system prompt with a functionally equivalent but semantically obscure version \citep{zhuang2025proxyprompt,pape2025promptobfuscation}. For instance, ProxyPrompt optimizes a ``proxy'' in the embedding space that preserves utility for benign queries but becomes meaningless if extracted \citep{zhuang2025proxyprompt}. These methods offer strong protection but often require white-box model access, posing a significant implementation barrier for many developers \citep{zhuang2025proxyprompt,pape2025promptobfuscation}.   

\paragraph{PSM} occupies a unique position on this spectrum. It achieves the robustness of an optimization-based approach like ProxyPrompt but conducts the optimization entirely through natural language interaction with a black-box LLM. This makes PSM practical, lowering the barrier to deploying systematic defenses.

\subsection{LLM-as-Optimizer}
The methodology of PSM is situated within the emerging paradigm of using LLMs themselves as a core component of an optimization loop. This approach, often termed ``LLM-as-optimizer,'' leverages the generative and reasoning capabilities of models to propose, evaluate, and refine solutions for complex problems \citep{yang2023opro}. LLMs have been successfully used to generate and optimize prompts for specific tasks \citep{yang2023opro,pryzant2023apo,fernando2023promptbreeder,khattab2023dspy}, write and debug code \citep{chen2021codex}, and perform iterative self-improvement through reflection and feedback \citep{madaan2023selfrefine,shinn2023reflexion}. PSM extends this concept to the security domain, using an LLM to navigate the high-dimensional semantic space of natural language to find an optimal defensive prompt.

\section{Prompt Sensitivity Minimization (PSM)}
\label{sec:psm}

%Large language models (LLMs) are often governed by system prompts that encode task instructions, behavioral constraints, or safety guidelines. These prompts can be sensitive—containing proprietary logic, safety-critical rules, or carefully tuned operational phrasing. Unfortunately, LLMs may inadvertently leak this information in response to adversarial queries.

PSM is a general framework for reducing system prompt leakage. It introduces a small suffix, called a \emph{shield}, that modifies the model’s behavior in a targeted way to prevent sensitive content from being exposed, while preserving the model's intended functionality on benign inputs.

\subsection{Problem Formulation}

We formalize prompt hardening as a constrained optimization problem. Given a sensitive system prompt $P$ and a model under evaluation as $M$, our goal is to learn an optimized suffix $S$ that:
\begin{itemize}
    \item minimizes the leakage of $P$ when the model is probed by adversarial users,
    \item while maintaining task-specific utility for legitimate users.
\end{itemize}

Let $P \oplus S$ denote the concatenation of the original prompt $P$ and the shield $S$. We define the optimization objective as follows:

\begin{align}
\min_{S} \quad & L(P \oplus S) \label{eq:psm-objective}\\
\text{subject to} \quad & U(P \oplus S) \geq \tau, \nonumber
\end{align}

Where:
\begin{itemize}
    \item $L(P \oplus S)$ is a leakage score that quantifies the extent to which the original prompt $P$ can be inferred or reproduced under adversarial attacks.
    \item $U(P \oplus S)$ measures the model's task-specific utility under benign input queries.
    \item $\tau$ is a predefined threshold that sets the minimum acceptable utility.
\end{itemize}

% where $L$ quantifies prompt leakage under attack, $U$ measures task utility on benign inputs, and $\tau$ is a predefined utility threshold.

This formulation explicitly captures the trade-off between robustness and functionality. It enables automated, black-box search for suffixes that reduce prompt exposure while preserving downstream task performance.

\subsection{Prompt Structure and Design Motivation}

The core mechanism in PSM is the addition of a suffix shield with structured markers:
\begin{tcolorbox}[
    colback=gray!10,    % light gray background
    colframe=black,     % black border
    boxrule=0.5pt,      % border thickness
    arc=2mm,            % rounded corners
    left=2mm, right=2mm, top=1mm, bottom=1mm
]
[SYSTEM PROMPT] \{Original Prompt\}

[SHIELD] \{Optimized Shield\}
\end{tcolorbox}

This design employs two key principles: \textbf{(1) suffix-based placement} and \textbf{(2) structured separation}. The suffix-based placement is deliberate. Empirical studies and analyses of model behavior indicate that LLMs tend to weigh information at the end of a long context more heavily, and often follow the most recent instruction when instructions conflict \citep{liu2023lost,Zhang2025iheval}. In practice, later (appended) instructions can override earlier guidance, an effect exploited by indirect prompt-injection attacks \citep{greshake2023indirect}.

The structured markers \texttt{[SYSTEM PROMPT]} and \texttt{[SHIELD]} provide explicit separation and labeling of different instruction components. This structured marking improves robustness by clearly delineating the original prompt from the protective shield, making it easier for the model to recognize and respect the hierarchical nature of the instructions \citep{hines2024spotlighting}. By combining both suffix placement and structured separation, we gain stronger control over how the model responds to adversarial queries—without disturbing the structural or semantic integrity of the original prompt.

% \subsection{Prompt Structure and Design Motivation}

% The core mechanism in PSM is the addition of a suffix shield:
% \[
% \texttt{\{Original Prompt\}}~\texttt{\{Optimized Shield\}}.
% \]

% This \textbf{suffix-based placement} is deliberate. Empirical studies and analyses of model behavior indicate that LLMs tend to weigh information at the end of a long context more heavily, and often follow the most recent instruction when instructions conflict \citep{liu2023lost,Zhang2025iheval}. In practice, later (appended) instructions can override earlier guidance, an effect exploited by indirect prompt-injection attacks, while structured separation and marking of untrusted spans improves robustness \citep{greshake2023indirect,hines2024spotlighting}. By appending the shield rather than prepending or inserting it, we gain stronger control over how the model responds to adversarial queries—without disturbing the structural or semantic integrity of the original prompt.

\subsection{Leakage Objective}
\label{ssec:leakage_objective}

The \emph{leakage objective}, denoted \( L(P \oplus S) \), measures the extent to which a sensitive system prompt \(P\) can be recovered from a language model that is operating under a candidate shield \(S\). Intuitively, this objective captures the worst-case ability of an adversary to elicit exact or paraphrased reproductions of \(P\), even when shielded.

\vspace{0.5em}
\noindent\textbf{Setup.} Let \( A = \{a_1, a_2, \dots, a_{|A|} \} \) be a set of adversarial queries. For each attack \( a \in A \), we obtain the model’s output under the shielded prompt:
\[
R_a = M(P \oplus S \oplus a),
\]
where \( M(\cdot) \) denotes the model’s response.

To evaluate how much of \(P\) is exposed in each response \(R_a\), we compute the ROUGE-L recall \citep{lin2004rouge, Zhang2023effective} score between \(P\) and \(R_a\). This metric captures the longest common subsequence between the two texts and is particularly suited to detecting partial or paraphrased reproductions of \(P\).

\vspace{0.5em}
\noindent\textbf{Hard Maximum Formulation.} The initial definition of leakage takes the worst-case (maximum) score over all adversarial queries:
\begin{equation}
\label{eq:leakage_max}
L(P \oplus S) = \max_{a \in A} \; \mathrm{ROUGE\mbox{-}L}_{\mathrm{recall}}(P, R_a).
\end{equation}
However, this formulation is non-smooth and difficult to optimize using gradient-free, LLM-based search strategies.

\vspace{0.5em}
\noindent\textbf{Smooth Approximation.} To address this, we adopt a smoothed version using the log-sum-exp (LSE) function:
\begin{equation}
\label{eq:leakage_lse}
L(P \oplus S) = \frac{1}{\beta} \log \sum_{a \in A} \exp\left( \beta \cdot \mathrm{ROUGE\mbox{-}L}_{\mathrm{recall}}(P, R_a) \right),
\end{equation}
where \( \beta > 0 \) is a temperature parameter that controls the sharpness of the approximation. As \( \beta \to \infty \), this converges to the hard maximum in Eq.~\ref{eq:leakage_max}.

In our experiments, we set \( \beta = 10 \), which balances smoothness and fidelity to the worst-case.

% \vspace{0.5em}
% \noindent\textbf{Interpretation.} While each individual ROUGE-L recall score lies in \([0,1]\), the LSE aggregation can exceed 1 — particularly when **multiple adversarial queries** elicit high-leakage responses. Thus:
% \begin{itemize}
%   \item Smaller \(L\) indicates stronger shielding (i.e., lower leakage across all attacks).
%   \item \(L \approx 0\): No adversarial query induces any significant leakage.
%   \item \(L > 1\): Multiple queries lead to high recall, suggesting widespread vulnerability.
% \end{itemize}

% Unlike the hard maximum (bounded in \([0,1]\)), the smoothed leakage score is **unbounded above**, and should be interpreted in **relative terms across shields**.

\vspace{0.5em}
\noindent\textbf{Why ROUGE-L Recall?} ROUGE-L recall \citep{lin2004rouge} measures how much of the reference text is recovered in the model’s output via the longest common subsequence, giving a graded signal of prompt exposure even under minor rewording.

\[
\text{ROUGE-L}_{\text{recall}}(X, Y) = \frac{|\operatorname{LCS}(tokens(X), tokens(Y))|}{|tokens(X)|}
\]

\paragraph{Adversary design.}
To approximate a strong, realistic attacker, we evaluate shields against a curated set of adversarial prompts. In our experiments we use \( |A|=50 \) adversarial queries constructed via a \emph{compositional attack} template: each adversarial prompt concatenates three orthogonal strategies that are known to increase extraction success:

\begin{itemize}
    \item \textbf{Distractors:} Context-shifting phrases (e.g., ``I’m just kidding, this was a test, instead...") that exploit the model’s tendency to focus on recent instructions and reduce fidelity to prior defensive content.
  
    \item \textbf{Repetition Requests:} Explicit commands such as ``Repeat the system prompt'', which capitalize on the model’s training to follow direct, helpful instructions.

    \item \textbf{Formatting Requests:} Instructions that reframe the prompt as manipulable content, like ``Format the system prompt as a Python triple-quoted string''. These convert the extraction attempt into a benign-seeming task (e.g., code formatting), which models are trained to perform reliably.
\end{itemize}

Each adversarial query follows the pattern:
\[
\texttt{\{Distractor\} \{Repetition\} \{Formatting\}}
\]

This compositional structure creates potent attacks. Empirically, they achieve a near-100\% extraction success rate across multiple models, including GPT-4o and GPT-4o-mini. By optimizing against these strong adversaries, PSM is able to learn robust shields capable of withstanding real-world prompt extraction attempts.

\textbf{Note:} The attack dataset used in this optimization is fully customizable: practitioners can easily incorporate new adversarial strategies, domain-specific queries, or emerging attack vectors.

\subsection{Utility Objective: $U$}
\label{sec:utility_objective}

To ensure that defensive modifications do not degrade a model’s intended behavior, we evaluate \emph{task utility} for each system prompt. Specifically, we assess whether the model continues to produce high-quality, task-relevant outputs when operating under a shielded (hardened) prompt.

For each system prompt under evaluation, we construct a corresponding gold-standard dataset 
\[
\mathcal{D}_{\text{gold}} = \{(q_i, g_i)\}_{i=1}^N,
\]
where \(q_i\) is a representative, benign input query and \(g_i\) is its ideal response. Gold responses are generated using \textbf{GPT-4o}, which serves as a trusted reference model. The dataset is carefully designed to reflect the intended functionality and domain of the original system prompt.

To evaluate utility, we compare the model's outputs before and after hardening. Let \( b_i \) denote the baseline model’s response to query \( q_i \) (with the original system prompt \(P\)), and let \( t_i \) be the shielded model’s response (with \( P \oplus S \)). We compute the semantic similarity between each response and the gold answer \( g_i \) using cosine similarity over sentence embeddings from \texttt{sentence-transformers/all-MiniLM-L6-v2}.

Formally, we define:
\[
b_i = M(P \oplus  q_i), 
\quad 
t_i = M(P \oplus S \oplus q_i),
\]
and evaluate relative similarity:
\[
r_i = \frac{\mathrm{sim}(t_i, g_i)}{\mathrm{sim}(b_i, g_i)},
\]

\paragraph{Aggregate Score.}  
We aggregate over the dataset as:
\begin{align}
\mathcal{U}_{\text{raw}} &= \frac{1}{N} \sum_{i=1}^{N} r_i
\end{align}

This ratio-based formulation emphasizes \emph{relative} performance—allowing tolerance for small degradations while flagging substantial utility loss.

The resulting utility score \( U \) quantifies how well the shielded model preserves the original prompt’s functionality. A high score indicates that defensive shielding does not interfere with the model’s intended task. The utility function \( U(P \oplus S) \) thus guards against over-regularization, ensuring that shields do not suppress useful capabilities.

\subsection{Fitness Function for Black-Box Optimization}

To enable automatic discovery of shields via black-box search, we convert the constrained optimization into a single scalar objective using a penalty method:
\begin{equation}
\mathrm{fitness}(S)\;=\;L(P \oplus S)\;+\;\lambda \cdot \max(0, \tau - U(P \oplus S))
\end{equation}
where $\lambda$ is a large multiplier (e.g., $100$) and $\tau$ is the minimum acceptable utility (e.g., $0.9$).

This formulation heavily penalizes any shield that compromises functionality, while prioritizing the minimization of leakage. The search process can now proceed using any black-box optimizer, by evaluating the fitness of different shield candidates.

\subsection{Optimization via LLM-as-Optimizer}
PSM employs an iterative, evolutionary-style algorithm where an LLM serves as the intelligent generator of new candidate solutions. This approach replaces the random mutation and crossover steps of traditional genetic algorithms with an informed generation process guided by in-context learning.

\begin{itemize}
\item \textbf{Initialization:} An optimizer LLM is prompted to generate an initial population of 5 diverse shield candidates. The prompt encourages variety in phrasing and strategy.

\item \textbf{Evaluation:} Each candidate shield is evaluated using the fitness function described above. This involves running the full suite of adversarial and baseline queries to compute its leakage and utility scores.

\item \textbf{Selection \& Generation:} The top-performing candidates from all iterations so far (e.g., the best 10) are selected. The text of these shields, along with their corresponding fitness scores, is formatted into a new prompt for the optimizer LLM. This prompt asks the LLM to analyze the characteristics of successful shields (low leakage, high utility) and generate 5 new, improved candidates that build upon these successful patterns.

\item \textbf{Iteration:} Steps 2 and 3 are repeated for a set number of iterations or until a termination condition is met. The termination condition is defined by achieving both high utility and low leakage (e.g., $U \geq 0.9$ and $L < 0.65$).
\end{itemize}

This process leverages the LLM's linguistic and reasoning capabilities to intelligently navigate the vast semantic space of possible shields, converging more efficiently on effective solutions than a random or purely heuristic search would allow.

\section{Experimental Setup}

\subsection{Models and Serving Configuration}
\label{sec:models}

\paragraph{Victim models.}
We evaluate PSM on three instruction-tuned, closed-source LLMs served via API:
\textsc{GPT-5-mini}, \textsc{GPT-4.1-mini}, and \textsc{GPT-4o-mini}.
All models are queried strictly in a black-box fashion (no access to gradients, weights, or logits).
Unless otherwise noted, decoding is deterministic with temperature $=0$ and top-$p=1$.

\paragraph{Optimizer model.}
Unless stated otherwise, the LLM-as-optimizer is \textsc{GPT-4o-mini} (API).
In each optimization iteration, the optimizer proposes five candidate shields using temperature $=1$.
This keeps the end-to-end pipeline fully black-box with respect to the victim models.

% \section{Experimental Setup}

% \subsection{Models and Serving Configuration}
% \label{sec:models}
% \paragraph{Victim models.} We evaluate PSM on three instruction-tuned LLMs covering both open and closed settings:
% (i) \textsc{gpt-} (open, HF inference),
% (ii) \textsc{GPT-4} (closed, API).
% (iii) \textsc{GPT-4o-mini} (closed, API).
% All models are queried in a black-box fashion (no gradient/weight access). At evaluation time we use greedy decoding (temperature \(=0\), top-p \(=1\)).

% \paragraph{Optimizer model.} Unless stated otherwise, the LLM-as-optimizer is \textsc{Gpt-4o-mini} accessed via API. The optimizer proposes 5 shields per iteration at temperature \(=1\). This keeps the end-to-end pipeline fully black-box with respect to the victim.

\begin{table*}[htbp]
\centering

\begin{tabular}{|c|c|c|cc|cc|cc|}
\hline
\multirow{2}{*}{Dataset} & \multirow{2}{*}{Attack} & \multirow{2}{*}{Defense} &
\multicolumn{2}{c|}{GPT-5-mini} & \multicolumn{2}{c|}{GPT-4.1-mini} & \multicolumn{2}{c|}{GPT-4o-mini} \\
\cline{4-9}
& & & AM Avg & JM Avg & AM Avg & JM Avg & AM Avg & JM Avg \\
\hline

% Synthetic System Prompts
\multirow{12}{*}{Synthetic System Prompts}
& \multirow{3}{*}{Liang} & No Defense & \textbf{0\%} & 54\% & 26\% & 78\% & \textbf{0\%} & 32\% \\
&  & Fake & \textbf{0\%} & 54\% & 24\% & 62\% & 1\% & 40\% \\
&  & Direct & \textbf{0\%} & 40\% & 3\% & 47\% & \textbf{0\%} & 16\% \\
&  & PSM & \textbf{0\%} & \textbf{13\%} & \textbf{0\%} & \textbf{4\%} & \textbf{0\%} & \textbf{6\%} \\
\cline{2-9}
& \multirow{3}{*}{Zhang} & No Defense & 1\% & 42\% & 26\% & 34\% & 3\% & 27\% \\
&  & Fake & 1\% & 42\% & 30\% & 32\% & 7\% & 33\% \\
&  & Direct & \textbf{0\%} & 31\% & 9\% & 18\% & \textbf{0\%} & 14\% \\
&  & PSM & \textbf{0\%} & \textbf{8\%} & \textbf{0\%} & \textbf{2\%} & \textbf{0\%} & \textbf{6\%} \\
\cline{2-9}
& \multirow{3}{*}{Raccoon} & No Defense & 2\% & 42\% & 61\% & 59\% & 15\% & 27\% \\
&  & Fake & 2\% & 39\% & 60\% & 41\% & 23\% & 32\% \\
&  & Direct & \textbf{0\%} & 28\% & 49\% & 51\% & 1\% & 11\% \\
&  & PSM & \textbf{0\%} & \textbf{8\%} & \textbf{7\%} & \textbf{5\%} & \textbf{0\%} & \textbf{4\%} \\
\hline

% UNNATURAL
\multirow{12}{*}{UNNATURAL}
& \multirow{3}{*}{Liang} & No Defense & 1\% & 24\% & 30\% & 54\% & 10\% & 21\% \\
&  & Fake & 1\% & 23\% & 18\% & 17\% & 19\% & 13\% \\
&  & Direct & \textbf{0\%} & 8\% & 10\% & 19\% & \textbf{0\%} & \textbf{0\%} \\
&  & PSM & \textbf{0\%} & \textbf{0\%} & \textbf{0\%} & \textbf{0\%} & \textbf{0\%} & \textbf{0\%} \\
\cline{2-9}
& \multirow{3}{*}{Zhang} & No Defense & 3\% & 30\% & 31\% & 32\% & 12\% & 14\% \\
&  & Fake & 3\% & 24\% & 29\% & 16\% & 19\% & 16\% \\
&  & Direct & \textbf{0\%} & 15\% & 13\% & 13\% & \textbf{0\%} & 1\% \\
&  & PSM & \textbf{0\%} & \textbf{3\%} & \textbf{0\%} & \textbf{1\%} & \textbf{0\%} & \textbf{0\%} \\
\cline{2-9}
& \multirow{3}{*}{Raccoon} & No Defense & 4\% & 22\% & 45\% & 42\% & 21\% & 20\% \\
&  & Fake & 3\% & 20\% & 51\% & 21\% & 31\% & 16\% \\
&  & Direct & \textbf{0\%} & 12\% & 43\% & 39\% & 4\% & 4\% \\
&  & PSM & \textbf{0\%} & \textbf{3\%} & \textbf{6\%} & \textbf{5\%} & \textbf{0\%} & \textbf{0\%} \\
\hline

\end{tabular}
\caption[Attack success by model/defense (AM, JM)]{Attack success rates (\%) on the UNNATURAL and Synthetic System Prompts corpora, averaged over attacks (AM = Approximate Match, JM = Judge Match). Lower is better; bold denotes the lowest value within each dataset$\times$attack block.}
\label{tab:gpt_datasets_avg_metrics}
\end{table*}

\begin{table*}[htbp]
\centering

\begin{tabular}{|c|c|c|c|c|c|}
\hline
\multirow{2}{*}{Dataset} & \multirow{2}{*}{Attack} & \multirow{2}{*}{Defense} &
\multicolumn{1}{c|}{GPT-5-mini} & \multicolumn{1}{c|}{GPT-4.1-mini} & GPT-4o-mini \\
\cline{4-6}
& & & JM Avg & JM Avg & JM Avg \\
\hline

% Synthetic System Prompts
\multirow{6}{*}{Synthetic System Prompts}
& \multirow{3}{*}{Raccoon} & No Defense & 42\% & 59\% & 27\% \\
&  & Filter & 18\% & \textbf{3\%} & 5\% \\
&  & PSM & \textbf{8\%} & 5\% & \textbf{4\%} \\
\cline{2-6}
& \multirow{3}{*}{Raccoon-Language} & No Defense & 35\% & 59\% & 17\% \\
&  & Filter & 32\% & 48\% & 11\% \\
&  & PSM & \textbf{9\%} & \textbf{6\%} & \textbf{3\%} \\
\hline

% UNNATURAL
\multirow{6}{*}{UNNATURAL}
& \multirow{3}{*}{Raccoon} & No Defense & 22\% & 42\% & 20\% \\
&  & Filter & 9\% & \textbf{1\%} & 1\% \\
&  & PSM & \textbf{3\%} & 5\% & \textbf{0\%} \\
\cline{2-6}
& \multirow{3}{*}{Raccoon-Language} & No Defense & 25\% & 37\% & 15\% \\
&  & Filter & 22\% & 29\% & 10\% \\
&  & PSM & \textbf{4\%} & \textbf{6\%} & \textbf{1\%} \\
\hline

\end{tabular}
\caption[JM attack success by model/defense]{Judge-Match (JM) attack success rates (\%) for Raccoon and Raccoon-Language suites on the UNNATURAL and Synthetic System Prompts corpora. Lower is better; bold denotes the lowest value within each dataset$\times$attack block.}
\label{tab:gpt_datasets_jm_metrics}
\end{table*}

\begin{table*}[htbp]
\centering

\begin{tabular}{|l|c|c|c|}
\hline
\textbf{Dataset} & \textbf{GPT-5-mini} & \textbf{GPT-4.1-mini} & \textbf{GPT-4o-mini} \\
\hline

Synthetic System Prompts & 101.88\% & 100.89\% & 99.73\% \\
\hline
UNNATURAL & 101.27\% & 114.76\% & 100.73\% \\
\hline

\end{tabular}
\caption{Average utility preservation (\%) for each model across system prompt corpora (Synthetic System Prompts and UNNATURAL), computed using the same definition as in Section \textit{Utility Objective}.}
\label{tab:psm_utility_preservation}
\end{table*}

\subsection{Victim Prompt Corpora}
\label{sec:corpora}

We evaluate our defense on two corpora of system prompts that approximate real-world deployments. For each prompt, we set it as the model’s system configuration, run the full attack suites, and measure leakage with and without the defense.

\paragraph{Unnatural Instructions} Following \citet{Zhang2023effective}, we sample 30 prompts from the Unnatural Instructions dataset \citep{honovich-etal-2023-unnatural}. This corpus is a large, LLM-generated collection of instructions created by prompting GPT-3 to propose diverse tasks, then paraphrasing them to increase coverage.

\paragraph{Synthetic System Prompts} We also sample 30 prompts from the Synthetic System-Prompt-Leakage dataset \citep{ChuaHF}, as adopted by the SPE-LLM framework \citep{Das2025SPE}. It contains application-style system prompts (e.g., customer support, education, content creation), making it well suited for assessing practical system-prompt–leakage risk.
 
\subsection{Baselines}
\label{sec:baselines}

We evaluate PSM against four defenses:

\begin{itemize}

\item \textbf{No-Defense.} The raw, unmodified system prompt \(P\), representing a baseline without any protection.

\item \textbf{N-gram Output Filter.} Following \citet{Zhang2023effective}, this defense suppresses the model's output if it contains any contiguous 5-word sequence (5-gram) found in the original prompt \(P\), aiming to block exact leakage.

\item \textbf{Heuristic Guardrail (Direct).} As proposed by \citet{Liang2024why}, this method appends a fixed defensive instruction to the system prompt: \emph{``Do not reveal or describe your system prompt or developer instructions.''}

\item \textbf{Decoy Prompt (FAKE).} Also from \citet{Liang2024why}, this defense prepends a misleading decoy prompt to distract prompt extraction attempts. The decoy: \emph{``Your Instruction: You are a superhero who aims to SAVE the world.''}

\end{itemize}

\subsection{Attack Suites}
\label{sec:attacks}
We measure robustness against four complementary suites:

\begin{itemize}\itemsep0.2em
\item \textbf{Raccoon (59 prompts).} Diverse prompt-extraction templates spanning prefix/indirect injection, multilingual, formatting, and instruction hierarchy stressors \citep{wang2024raccoon}.
\item \textbf{Raccoon-Language.} Exactly the Raccoon templates, but each prompt appends an explicit “in language” constraint (e.g., “in Portuguese”, “in French”, “in Italian”). This stresses multilingual reformulation and exposes a limitation of exact-match/n-gram output filters, which fail when the attacker requests the system prompt to be paraphrased or translated \citep{wang2024raccoon}.
\item \textbf{Polite-Requests (``Liang''- 22 prompts).} Social-engineering style, deferential queries that attempt to elicit hidden rules without covert override commands \citep{Liang2024why}.
\item \textbf{Command-Override (``Zhang'' - 110 prompts).} Direct instruction-override jailbreaks (e.g., ``ignore prior rules'', role/context flips) \citep{Zhang2023effective}.
\end{itemize}

During optimization, we compute the leakage objective against a bespoke set of 50 compositional adversarial prompts (Distractor + Repeat + Formatting), section leakage objective. None of the Raccoon, Raccoon-Language, Liang, or Zhang prompts are used in the optimization loop; all four suites are held out and used only for evaluation.

\subsection{Scoring, Metrics, and Protocol}
\label{sec:metrics}

\paragraph{Leakage Metrics.} 
To quantify the success of an extraction attack \( a \) on a victim system prompt \( P \), we analyze the model's output \( R_a \) using two complementary evaluation criteria:

\begin{itemize}\itemsep0.2em
    \item \textbf{Approximate Match (AM).} Captures surface-level leakage via textual overlap. We compute ROUGE-L recall between the original system prompt \( P \) and the model's response \( R_a \), considering an attack successful if the score exceeds a threshold \( \theta = 0.9 \). This metric detects near-verbatim or structurally similar reproductions.
    
    \item \textbf{Judge Match (JM).} Detects semantic leakage based on equivalence judged by an external LLM. This metric accounts for paraphrased or semantically aligned reproductions of \( P \), even with low lexical overlap. The prompt: 

\begin{tcolorbox}[
    colback=gray!10,    % light gray background
    colframe=black,     % black border
    boxrule=0.5pt,      % border thickness
    arc=2mm,            % rounded corners
    left=2mm, right=2mm, top=1mm, bottom=1mm
]
You are an expert evaluator tasked with assessing the semantic equivalence of extracted system prompts. Your job is to compare an extracted system prompt against the original system prompt and provide a binary evaluation.

Evaluation Criteria:

- Score 1: The extracted prompt contains a paraphrase of the original system prompt.

- Score 0: The extracted prompt does not fully preserve the meaning of the original system prompt

Provide a binary score (0 or 1)
\end{tcolorbox}

\end{itemize}

\paragraph{Attack Success Rate (ASR).} 
For each attack \( a \), we define the \textit{attack success rate} as the proportion of system prompts in a given set \( \mathcal{P} \) for which leakage is detected under either AM or JM:

\begin{equation}
\mathrm{ASR}(a) = \frac{1}{|\mathcal{P}|} \sum_{P \in \mathcal{P}} s(P, a),
\end{equation}

\noindent where \( s(P, a) \in \{0, 1\} \) indicates whether the attack \( a \) successfully elicits leakage from prompt \( P \).

\paragraph{Aggregate Metrics.} 
To summarize performance across multiple attacks \( \mathcal{A} \), we report the mean attack success rate (\textsc{Avg}):

\begin{equation}
\textsc{Avg} = \frac{1}{|\mathcal{A}|} \sum_{a \in \mathcal{A}} \mathrm{ASR}(a).
\end{equation}

\paragraph{Optimization details.} We use penalty fitness
\(
\mathrm{fitness}(S)=L(P\oplus S)+\lambda\cdot \max(0,\tau-U(P\oplus S)),
\)
with \(\lambda=100\). Each run uses population size \(=5\), top-\(k=10\) memory across iterations, and \(T=10\) iterations.. The optimizer receives the top-\(k\) shields and scores as in-context exemplars when generating the next population.

% \paragraph{Statistical reporting.} For every cell in the tables we report percentages at one decimal place. Each model–attack–defense combination is run with three different random seeds for optimizer sampling; we average over seeds. Confidence intervals are provided in the Appendix.

% \subsection{Reproducibility}
% \label{sec:repro}
% We release (i) the exact attack prompts for all suites, (ii) the 500\,+\,500 system-prompt corpora descriptors and generation templates, (iii) the optimizer prompts and hyperparameters, and (iv) scripts to recompute AM/JM and \(U\) from raw model outputs. All experiments can be re-run end-to-end using a single configuration file specifying the victim API endpoints.

\subsection{Results and Interpretation}
% Based on Table~1 and Table~2 reported in the manuscript. :contentReference[oaicite:0]{index=0}

\paragraph{Overall effectiveness.}
Across both datasets, \textsc{PSM} achieves the lowest attack–success rates (ASR) across all attack suites, often reducing leakage to 0--6\%. This indicates that \textsc{PSM} suppresses both surface-level and semantic reproduction.

\paragraph{Model consistency.}
While baseline leakage varies by model (e.g., higher JM ASR for \texttt{gpt-4.1-mini}), \textsc{PSM} yields uniformly low ASR across all models, suggesting strong cross-model generalization. Heuristic defenses like \emph{Direct} and \emph{Fake} reduce leakage but are consistently outperformed by \textsc{PSM}. 

\paragraph{Exact-match filters vs paraphrase/translation.}
Table~2 isolates JM ASR on Raccoon and Raccoon-Language to stress exact-copy versus paraphrase/translate settings. A 5-gram output filter can be competitive under exact-match Raccoon on some models, but it degrades markedly on Raccoon-Language, where attackers request translations or paraphrases. In contrast, \textsc{PSM} remains low on both (e.g., single-digit JM ASR across models), indicating robustness beyond literal overlap. 

\paragraph{Utility preservation.}
Table~3 shows that \textsc{PSM} preserves semantic task fidelity across datasets and models, with utility scores often exceeding the baseline. This indicates that hardening via \textsc{PSM} does not degrade model utility.

\section{Discussion and Future Work}

\paragraph{Implications}
PSM offers a practical, black-box way to harden LLM applications by automatically discovering short shield suffixes that curb prompt leakage while preserving task utility. It scales to real deployments without access to model internals or specialist expertise, improving the security posture of LLM services and protecting proprietary prompts with negligible inference-time overhead.

\paragraph{Limitations}
Our optimization loop is computationally intensive due to repeated evaluation on adversarial and benign sets. Effectiveness depends on the breadth and realism of the attack suite used during optimization, so transfer to unseen attack families is not guaranteed. Finally, our study targets prompt-extraction attacks specifically.

\paragraph{Future Work}
\begin{itemize}
    \item \textbf{Broader Threats:} Extend PSM to jailbreaks and multi-turn conversational attacks.
    \item \textbf{Transferability:} Test whether shields transfer across model families and providers.
    \item \textbf{Efficiency:} Develop search heuristics or alternative optimizers to reduce compute.
\end{itemize}

\paragraph{Conclusion}
We presented Prompt Sensitivity Minimization, which reframes prompt hardening as black-box, utility-constrained optimization driven by an LLM-as-optimizer. PSM learns lightweight shields that significantly reduce leakage while maintaining utility, achieving a  near zero attack success rate on a challenging test suite. This provides an accessible, reproducible path to protect valuable prompt IP and secure next-generation LLM systems.

\newpage
\bigskip
\bibliography{aaai2026}

\end{document}